\documentclass{article}
\usepackage[english]{babel}
\usepackage{authblk}
\usepackage[letterpaper,top=2cm,bottom=2cm,left=2cm,right=2cm,marginparwidth=1.75cm]{geometry}
\usepackage{adjustbox}
\usepackage{float}

\usepackage{graphicx}
 \usepackage{textcomp}
\usepackage[colorlinks=true, allcolors=blue]{hyperref}
\usepackage{amsmath}
\makeatletter
\let\@seccntformat\@gobble
\makeatother

\begin{document}
\date{} 
\title{Inkjet-Printed High-Yield, Reconfigurable, and Recyclable Memristors on Paper}

\author{Jinrui Chen\textsuperscript{†,1}, Mingfei Xiao\textsuperscript{†,1}, Zesheng Chen\textsuperscript{1}, Sibghah Khan\textsuperscript{1},\\
Saptarsi Ghosh\textsuperscript{2}, Nasiruddin Macadam\textsuperscript{1}, Zhuo Chen\textsuperscript{1}, Binghan Zhou\textsuperscript{1},\\
 Guolin Yun\textsuperscript{1}, Kasia Wilk\textsuperscript{2}, Feng Tian\textsuperscript{1,3}, Simon Fairclough\textsuperscript{2}, Yang Xu\textsuperscript{3}, Rachel Oliver\textsuperscript{2},\\Tawfique Hasan\textsuperscript{*,}}

\affil[1]{Cambridge Graphene Centre, University of Cambridge, Cambridge, CB3 0FA, UK}
\affil[2]{Department of Materials Science \& Metallurgy, University of Cambridge, Cambridge, CB3 0FS, UK}
\affil[3]{School of Micro-Nano Electronics, ZJU-Hangzhou Global Scientific and Technological Innovation Center, State Key Laboratory of Silicon and Advanced Semiconductor Materials, ZJU-UIUC Joint Institute, Zhejiang University, Hangzhou 310027, China}
\renewcommand*{\Affilfont}{\small\it}

\maketitle
\begin {abstract}
\noindent
Reconfigurable memristors featuring neural and synaptic functions hold great potential for neuromorphic circuits by simplifying system architecture, cutting power consumption, and boosting computational efficiency.
Their additive manufacturing on sustainable substrates offers unique advantages for future electronics, including low environmental impact.
Here, exploiting structure-property relationship of MoS$_2$ nanoflake-based resistive layer, we present paper-based, inkjet-printed, reconfigurable memristors.
With \textgreater90\% yield from a $16\times65$ device array, our memristors demonstrate robust resistive switching, with \textgreater$10^5$ ON-OFF ratio and \textless0.5 V operation in non-volatile state.
Through modulation of compliance current, the devices transition into volatile state, with only 50 pW switching power consumption, rivalling state-of-the-art metal oxide-based counterparts.
We show device recyclability and stable, reconfigurable operation following disassembly, material collection and re-fabrication.
We further demonstrate synaptic plasticity and neuronal leaky integrate-and-fire functionality, with disposable applications in smart packaging and simulated medical image diagnostics.
Our work shows a sustainable pathway towards printable, high-yield, reconfigurable neuromorphic devices, with minimal environmental footprint.
\end{abstract}

\section{Main}
Neuromorphic computing systems, exemplified by memristors, have immense potential in simplifying computational and storage structures with improved computational efficiency.\cite{ref1, ref2, ref3}
Advancements in neuromorphic engineering not only offer a new technological pathway for future cognitive systems on silicon,\cite{ref4,ref5,ref6,ref7,ref8} but also align with the demands of organic electronics for low-power consumption and high integration potential.\cite{ref9,ref10,ref11,ref12}
However, challenges in fabricating these versatile neuromorphic systems often arise from the distinct materials and manufacturing techniques required for various components, such as electronic synapses and neurons. 
These variations can introduce technical complexity, increase resource consumption, and lead to material incompatibility challenges.\cite{ref13, ref14}
In addition, traditional semiconductor- and plastic-based commercialised neuromorphic devices may also contribute to the rapidly growing e-waste problem.\cite{ref15}
From this perspective, additive manufacturing of reconfigurable memristors on sustainable substrates offers an efficient and eco-friendly approach for neuromorphic systems in transient or disposable applications.
Reconfigurable memristors, with their ability to emulate both synaptic and neuronal functions, greatly simplify the circuitry architecture and enhance integration density.\cite{ref16, ref17} 
Although there have been some efforts in additive manufacturing of simple, traditional electronic circuits on paper,\cite{ref18} exploration of multifunctional reconfigurable memristors on this eco-friendly substrate has remained unexplored. 

Here we demonstrate inkjet-printed reconfigurable memristors on paper. 
Achieving a \textgreater90\% production yield from a 16 $\times$ 65 array, the  memristor devices exhibit an ON-OFF ratio of up to $10^5$ and $< 0.5$ V operation voltage. 
Due to the unique stacking configuration created in inkjet printed layer and abundant sulfur vacancies generated in MoS$_2$ during exfoliation, the Ag/MoS$_2$/Au-structured devices exhibit both non-volatile resistive switching (RS) and volatile switching properties, allowing synaptic and neuronal functionalities in a single device. 
We demonstrate that the majority of device components can be recycled and reused, with re-fabricated devices also exhibiting robust performance, with an ON-OFF ratio exceeding $10^5$ and $< 0.65$ V operational voltage. 
To illustrate the versatile application scopes of our memristors, we first explore the reconfigurability of individual devices and demonstrate tri-mode electronic time-temperature indicators (TTI) on a paper substrate for smart packaging application.
Subsequently, we extend our focus to the system level, where we develop a lightweight artificial neural network (ANN)-based algorithm for efficient extraction of key features from color images. 
We then adapt this algorithm for diabetic retinopathy (DR) screening. 
Accounting for device variation extracted from our 16$\times$65 printed memristor array, simulation results indicate that this system identifies all four types of lesions, achieving \textgreater90\% specificity and accuracy indices.
Our paper-based reconfigurable memristors show clear advantages of multi-functionality and low-cost, high-yield manufacturability at a large scale, and hold promise towards future sustainable electronics.

\section{Nanoflake ink formulation and inkjet printing}
Figure 1a illustrates the process, starting from ink formulation to device fabrication and recycling pathways. 
The high-pressure homogenization (HPH) process exfoliates bulk MoS$_2$ powder (Sigma-Aldrich) into few-layer nanoflakes in isopropyl alcohol (IPA).\cite{ref19}
This procedure leverages the combined effects of cavitation, shearing, and impact forces within the shearing chamber. 
By controlling the number of processing cycles, we can tune the lateral dimensions of the exfoliated 2D materials while maintaining a relatively consistent thickness.
Atomic force microscopy-based (AFM) statistics in Figure S1 reveals that the lateral size of the exfoliated MoS$_2$ flakes after 250 cycles is 42$\pm$15 nm with a thickness of 2.6$\pm$0.8 nm.
The uniformly sized nanoflakes ensure a stable inkjet printing process, crucial for maintaining consistency in large-area device fabrication.
This uniformity also promotes a homogeneous surface topology, pivotal in ensuring high device yield and reducing device-to-device variation.\cite{ref20} 
UV-Vis absorption spectroscopy of the MoS$_2$-IPA dispersion with varying diluted concentrations exhibits two characteristic excitonic peaks (Fig. 1b), located near 670 nm (A) and 615 nm (B). 
These peaks correspond to the direct bandgap excitonic transitions at the \textit{K} point in the Brillouin zone of the 2H phase of MoS$_2$, indicating that the inherent phase of MoS$_2$ has been preserved during the exfoliation process.\cite{ref21, ref22} 
As shown in Fig. 1c, X-ray photoelectron spectroscopy (XPS) on MoS$_2$ nanoflakes shows characteristic peaks at 229 eV (Mo 3d$_{5/2}$), 232.2 eV (Mo 3d$_{3/2}$), 163 eV (S 2p$_{1/2}$), and 161.8 eV (S 2p$_{3/2}$), respectively.
These detected binding energies align with those expected for pristine MoS$_2$ without oxidation.\cite{ref22} 
The XPS results also reveal the stoichiometry of Mo/S with a ratio of 1:1.83, indicating the generation of sulfur vacancies during the HPH process.
We hypothesise that the vacancies foster interlayer diffusion of conductive filaments (CF), engendering efficient pathways to modulate RS, similar to the observation for grain-boundaries in hexagonal boron nitride.\cite{ref23} 
Therefore, the uniform nanoflakes with sulfur vacancies exfoliated via the HPH technique serve as a robust basis for the fabrication of filamentary memristors.

The exfoliated flakes are stably dispersed in a binary solvent of IPA/2-Butanol (10 volume\%) to formulate the ink.\cite{ref24, ref25} 
The resultant ink allows smooth inkjet printing without nozzle clogging.
The use of binary solvents also helps suppressing non-uniform deposition.\cite{ref24} 
Figure 1d illustrates an example of the inkjet-printed pattern produced using the formulated ink on paper substrates. 
The insets showcase the corresponding optical micrographs of selected areas within Fig. 1d, illustrating the uniform deposition and clear edges of the printed patterns. 
We define printing accuracy in terms of pattern fidelity, which quantitatively assesses the alignment accuracy between the designed and the actual printed patterns.
It is assessed by overlaying the printed and imported images, and quantifying the deviation as a percentage of the total printed area (Fig. S2).
At the microscopic scale (scale bar 200 \textmu m), the average printing accuracy reaches 98\%. 
Stable jetting and such high printing accuracy with our ink formulation enable uniform material deposition for device fabrication.

\begin{figure}[H]
\centering
\includegraphics[width=\textwidth]{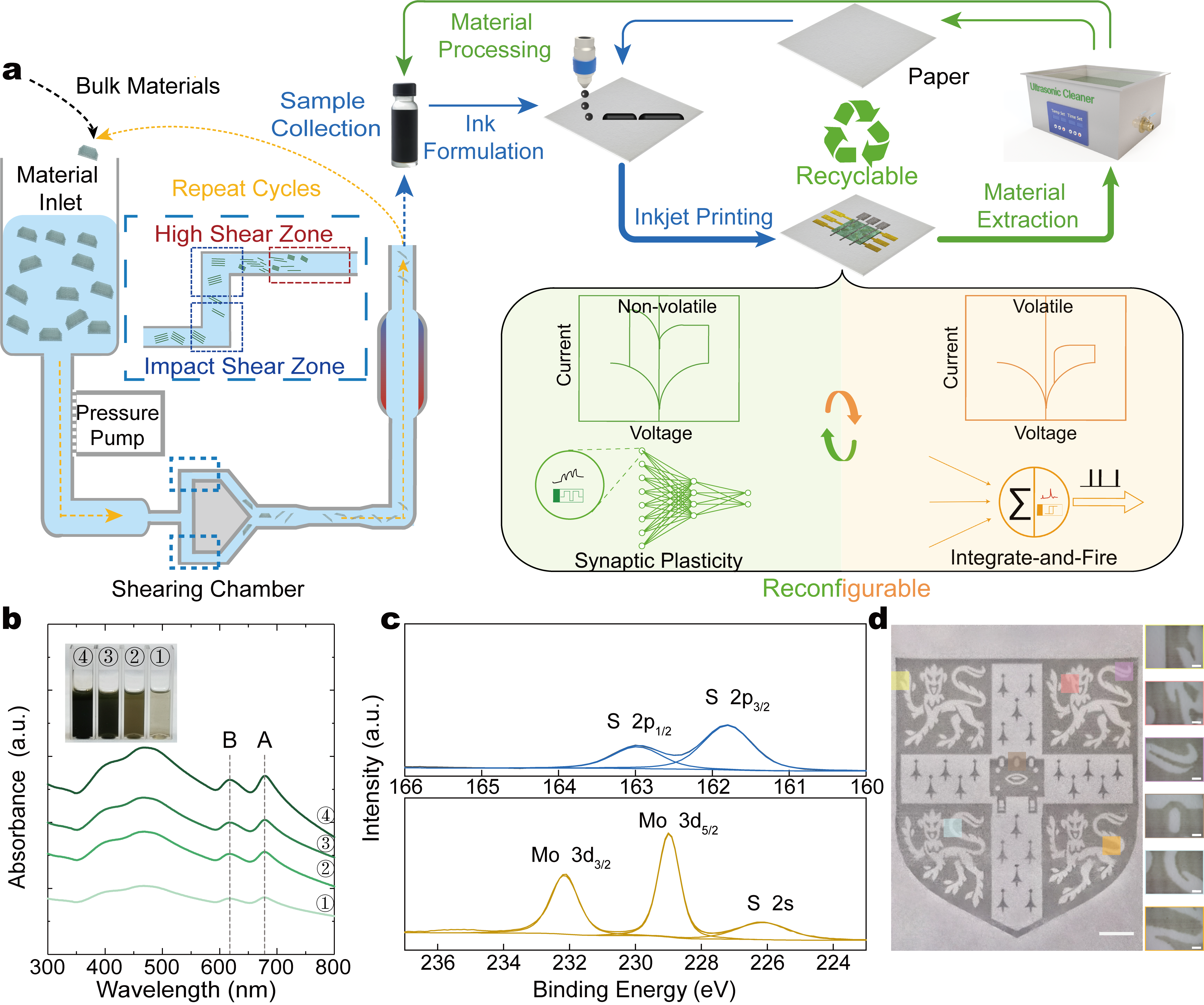}
\caption{\label{fig:1}
\textbf{Schematic of the HPH process and characterization of the MoS$_2$ ink.}
\textbf{a} Schematic of the ink preparation, device fabrication, performance and reconfiguration and the recycling pathways. 
\textbf{b} UV-Vis absorption spectra of the MoS$_2$-IPA dispersion at different concentrations. The inset displays a photograph of the corresponding MoS$_2$-IPA dispersions. 
\textbf{c} XPS characterization of the exfoliated MoS$_2$ sample. 
\textbf{d} Inkjet printed Coat of Arms of the University of Cambridge with the MoS$_2$ ink on PEL P60 paper substrate. The scale bar represents 0.5 cm. The insets are optical micrographs of the highlighted areas in larger printed pattern, showing high print quality. The scale bars in the insets represent 250 \textmu m.
}
\end{figure}

\section{Characterization of inkjet-printed 2D memristors}
The printed memristors, featuring an Ag/MoS$_2$/Au structure, are fabricated on PEL-60 paper substrate (See method). 
The high solvent absorption capacity of the substrate facilitates rapid ink drying after deposition.
In addition to the use of binary solvents, this further aids suppression of non-uniform deposition and avoids the need for high-temperature annealing to remove the carrier solvents trapped in the printed layers.

To investigate the impact of active layer thickness on RS characteristics in the printed devices, we fabricate an array of 5,200 ($80\times65$) memristors; Fig. 2a.
The array is divided into five sections, with increasing number (from 10 to 50 print repetitions) of MoS$_2$ layers.
Figure 2b shows the top-right corner of the array while Fig. 2c displays a $3\times4$ set of devices within this region. 
For each $16\times65$ section, 100 devices are randomly selected for measurement (Fig. S3).
Statistical analysis reveals that 30-layer devices exhibit the highest yield of 93\% (Fig. 2d, S4). 
Moreover, 92.5\% of the working devices exceed an ON-OFF ratio of 5 decades (Fig. 2f), with a mean ON-OFF ratio of $5.5\pm0.3$ decades (Fig. S4), indicating well-managed device-to-device variation (Fig. S5). 
The 30-layer device also strikes an optimal balance between the ON-OFF ratio and operation voltage (Fig. S6).
Therefore, the 30-layer device is used in all subsequent evaluations unless stated otherwise. 
Cross-sectional transmission electron microscopy (TEM) (Fig. 2e) and the corresponding energy-dispersive X-ray spectroscopy (EDS) elemental mapping (Fig. S7) reveal that the 30-layer device has an active layer thickness exceeding 500 nm with minimal oxidation.
This may result in a high initial resistance beneficial for achieving a large ON-OFF ratio.
It also reveals that upon deposition via inkjet printing, the exfoliated nanoflakes create a stacked architecture.
This introduces a plethora of potential diffusion pathways.
These pathways are essential for the formation of CFs and lay the foundation for low voltage operation of the device. \cite{ref26} 

Under different compliance current (CC) conditions, we can control the formation and rupture of CFs to manipulate the transition between volatile and non-volatile states. 
When the CC is 1 mA, the application of SET voltage (positive voltage) oxidizes the silver atoms into silver ions, which migrate through the MoS$_2$ layer and connect to the grounded bottom electrode, forming robust CFs.\cite{ref1, ref27} 
These filaments remain stable for more than 5000 s (Fig. S8), demonstrating excellent data retention capabilities.
Thus, the device necessitates a RESET process to break these filaments using negative voltage (Fig. 2g), which might be a consequence of the synergistic effects of both the dissolution of Ag through reverse ionic diffusion and thermally induced filament disruption. 
Under this CC, the ON-OFF ratio of the memristor reaches $10^5$ with a sharp turn-on slope of $3.07$ mV decade$^{-1}$, accompanied by mean SET voltage of 0.29$\pm$0.10 V and RESET voltage of -0.44$\pm$0.20 V (Fig. S9).  
Notably, the critical RS performance parameters outlined above, such as operational voltage below 0.5 V and ON-OFF ratio exceeding $10^5$, compare favorably with those of other high-performance memristors fabricated by chemical vapor deposition (CVD) or complementary metal-oxide semiconductor (CMOS) techniques; see Table S1.
When the CC is decreased to 0.01 mA, the devices enter the volatile state (Fig. 2h). 
In this regime, thinner CFs formed during the SET process tend to spontaneously dissociate upon removal of the electric field due to thermally assisted diffusion and minimum energy effect.\cite{ref29, ref30, ref31} 
This exemplifies typical volatile switching behavior. 
With an ON-OFF ratio of $10^5$, the average SET voltage of the memristor is $0.40\pm0.19$ V (Fig. S10), with a corresponding power consumption of only 50 pW. 
The high ON-OFF ratio and low power consumption of the volatile switching make it ideal for use as selectors to strongly reduce sneak currents within crossbar arrays of memory devices.\cite{ref31}

Subsequently, to elucidate the operating mechanisms of our devices, we deploy a range of complementary techniques, including conduction mechanism analysis, electrode substitution, in-situ microscopic examination and defect engineering. 
Conduction mechanism analysis indicates that the formation, rupture, and self-dissolution of CFs significantly affect the operational mechanism of the device (see Fig. S11 for detailed notes).
Then, we carry out electrode substitution experiments to confirm that the Ag electrodes dominate these dynamics of CFs.
We use inkjet-printed graphene (Gr) electrodes to substitute either the top Ag or the bottom Au electrodes.
The Gr/MoS$_2$/Au device, without top Ag electrode, shows no RS performance (Fig. S12a).
Conversely, the Ag/MoS$_2$/Gr device, with the bottom electrode switched from Au to Gr, exhibits volatile RS with a reduced operation voltage of \(0.30 \pm 0.10\) V (Fig. S12b) compared to the original devices (Fig. 2h).
The volatile RS performance can be attributed to the chemically inert characteristic of graphene against Ag atoms, which hinders the stability of the formed Ag CFs to facilitate volatile switching.\cite{ref32, ref33}
This implies possible future opportunities to further reduce energy consumption and fine-tune the device operational voltage by simply replacing electrode materials.

Next, we use conductive AFM (CAFM) in contact mode to investigate the dynamics of Ag CFs at nanoscale (Fig. S13a). 
We replace the bottom Au electrode with the Pt-Ir CAFM probe.
With the Pt-Ir probe grounded and a positive voltage sweep applied on the inkjet-printed silver, the Ag/MoS$_2$/Pt-Ir structure behaves like a nanoscale memristor.
In the first sweep, we observe a volatile RS switching with 10$^4$ ratio. 
With repeated voltage scanning, the SET voltage and ON-OFF ratio gradually decrease, and the device eventually stabilizes in the LRS (Fig. 2j). 
This conductance increase implies synaptic plasticity, providing in-situ evidence of our device’s transition from volatile state to non-volatile state.\cite{ref34, ref35} 
Upon applying a higher voltage to the sample, the location where thick and persistent filaments form, can be directly observed in the current map (Fig. S13b,c).

Lastly, we study the influence of sulfur vacancies on RS performance of our devices. 
We fabricate MoS$_2$ memristor devices with 10 to 40 printed layers (Fig. S14) to serve as a control group in comparison to the original devices.
Following this, we employ dithiolated conjugated molecule 1,4-benzenedithiol (BDT) to heal the sulfur vacancies and to facilitate the covalent bridging of the neighbouring MoS$_2$ flakes (see Methods).\cite{ref36}
Due to the healing of vacancies, all the BDT-treated devices (Fig. S14) show a larger breakdown voltage of MoS$_2$ layer compared to the untreated counterparts (Fig. S6). 
For example, the 30-layer device requires a forming voltage of ~ 9 V (Fig. 2k), while its untreated counterpart operates at \textless 0.5 V (Fig. 2g). 
Upon conducting subsequent I-V sweep operations, all the BDT-treated devices display a noticeable decrease in the ON-OFF ratio, eventually leading to the complete disappearance of the RS phenomenon within three scans or fewer (see Fig. 2k, S14 for 10-40 layers of the BDT-treated devices). 
We attribute these phenomena to the inhibition of silver ion permeation pathways within the MoS$_2$ layers caused by the healing of vacancies through BDT treatment.
This inhibition may impede the formation of CFs in MoS$_2$.
Only a high forming voltage can breakdown the MoS$_2$ layer and form the CFs to set the device to LRS.
However, the filaments formed under such excessively high voltage may cause irreversible damage to the devices, resulting in a minimal ON-OFF ratio. 
The retention measurements indicate that even when the CC is below 0.01 mA, the BDT-devices maintain the stored information, unlike the original devices that transition into a volatile state under same CC (Fig. S14d).
The sulfur vacancies may promote the dissociation of Ag CFs by enhancing their instability under low CCs. 
Thus the healing of these vacancies may extend the retention of CFs.
Crucially, our results highlight that the sulfur vacancies introduced by the HPH process and the permeation pathways within MoS$_2$ layer may promote the RS and reconfigurable functionality of the device. 

Recyclable electronics are highly desirable in reducing the ever increasing environmental footprint of electronic waste and promoting sustainable development. 
This is particularly important for transient or disposable applications.\cite{ref37}
In light of this, we explore the recyclability of our memristors. 
Our device exhibits an endurance of over 2500 cycles (Fig. S15). 
Upon reaching the end of its lifespan, all components of the printed memristor can be recycled and reused. 
In particular, we can detach the inkjet-printed MoS$_2$ and the silver electrodes on MoS$_2$ via simple ultrasonic cleaning (Fig. S16a,b). 
The detached MoS$_2$ nanoflakes and Ag nanoparticles are recollected and separated (Fig. S16c) through vacuum filtration and high-speed centrifugation (See method). 
This allows re-fabrication of the MoS$_2$ layer and silver electrodes onto the same paper substrate with the gold electrode (Fig. S16d). 
Consistent with the performance of the original device, the reprinted memristor under different CCs demonstrates a transition between non-volatile (Fig. S16e) to volatile state (Fig. S16g). 
Despite variations in the operational voltage (Fig. S16f,h), the reprinted device still exhibits an ON-OFF ratio exceeding $10^5$ and mean operational voltages below 0.65 V (Fig. S16e,g), affirming its consistent reliability throughout the recycling process. 

\begin{figure}[H]
\centering
\includegraphics[width=\textwidth]{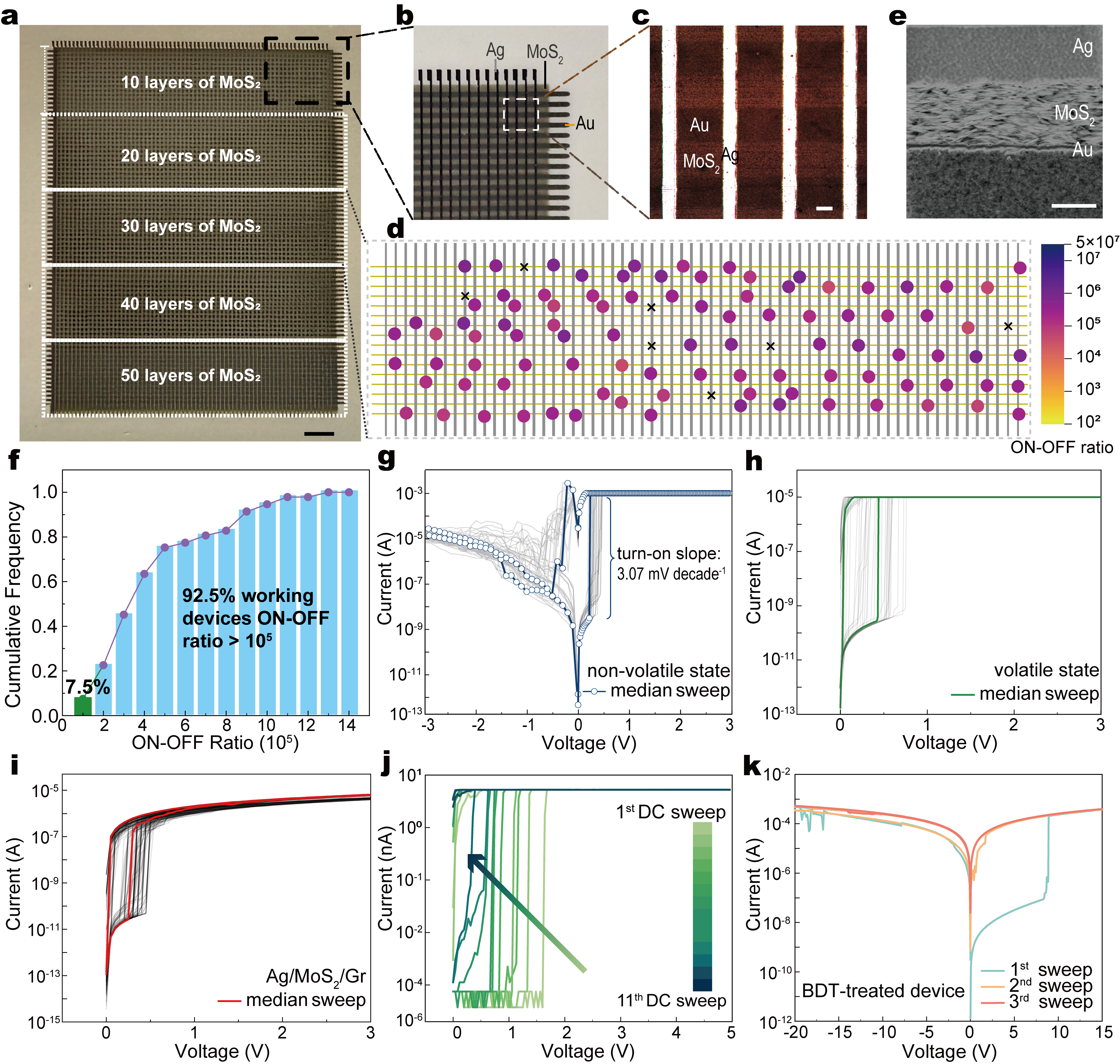}
\caption{\label{fig:2}
\textbf{Electrical property characterization and working mechanism investigation of inkjet-printed memristors.}
\textbf{a} Photograph of a $80\times65$ inkjet-printed memristor array on paper with 10, 20, 30, 40 and 50 printed MoS$_2$ layers, with $16\times65$ devices for each thickness. The scale bar represents 2 mm. 
\textbf{b} Close-up of the area enclosed by the dashed line in \textbf{a}. 
\textbf{c} Optical micrograph of a $3\times4$ set of devices from the area enclosed by the dashed line in \textbf{b}. The scale bar represents 50 \textmu m. 
\textbf{d} Cross-sectional TEM image of a 30-layer device. The scale bar is 500 nm. 
\textbf{e} Performance of randomly selected 100 devices in the 30-layer $80\times65$ memristor array.
The position of the dots represents their location within the array while their color indicates ON-OFF ratios. The crosses represent devices that do not exhibit RS behavior.   
\textbf{f} Cumulative frequency of the ON-OFF ratio for 93 functional devices out of 100 randomly selected, 30-layer devices. 
Among the functional devices, those with an ON-OFF ratio of $\textgreater 10^5$ are highlighted by the blue bars.
\textbf{g} 50 non-volatile resistive switching operations of a selected memristor under a CC of 1 mA, the median sweep is highlighted. 
\textbf{h} 50 volatile resistive switching operations (in the same device as \textbf{g}) under a CC of 10 \textmu A, the median sweep is highlighted. 
\textbf{j} CAFM characterization of the Ag/MoS$_2$ sample. 11 voltage sweeps ($0\text{V}-5\text{V}-0\text{V}$) are performed at the same location on the inkjet-printed MoS$_2$. The results demonstrate that with successive voltage sweeps, both the ON-OFF ratio and the SET voltage of the sample gradually decrease.
\textbf{k} Resistive switching performance of the BDT-treated memristor device with 30 printed layers, revealing a higher SET voltage and a significantly reduced ON-OFF ratio in subsequent scans, compared to the untreated device (Fig. 2g).}
\end{figure}

\section{Reconfigurable memristive performance in inkjet printed devices}

Due to the ability to dynamically transition between non-volatile and volatile states, our reconfigurable memristors can effectively emulate both synaptic and neuronal functionalities. 
This emulation is achieved by varying voltage amplitudes and duration: lower voltage pulses enable transient changes akin to neuronal firing, while higher voltages induce synaptic-like behavior characterized by gradual resistance changes after the CFs form, mimicking synaptic plasticity.

Neurons process information by integrating incoming signals (Fig. 3a). 
When subjected to external stimuli, the neuron, initially at a resting potential, begins to integrate voltage changes from these input signals into its membrane potential.\cite{ref38, ref39}
Once the membrane potential reaches a specific threshold, the neuron `fires' an action potential.
The membrane potential swiftly resets to the initial resting state, preparing to respond to subsequent stimuli.\cite{ref40}
This leaky integrate-and-fire (LIF) functionality can be replicated by memristors operating in a volatile threshold switching state (Fig. 3b). 
In response to a continuous voltage pulse train (100 \textmu s, 1 V; 100 \textmu s, 0.1 V), no current response is detected until a sudden surge after receiving the 5$^{th}$ pulse, mimicking the integrate-and-fire behaviour. 
The small current spikes occurring at the edge of voltage pulses are possibly induced by parasitic capacitance. 
After firing, the current falls to the limit of detection again in the subsequent voltage pulses, indicating that the memristor has recovered to the low conductance state and is prepared for the next-cycle integration process. 
This self-recovery stems from the fragile nature of the Ag CFs. 
We suggest that the formed CFs are thin and prone to spontaneous dissociation, which emulates the repolarization behavior of neurons.\cite{ref33}
Figure 3c illustrates the relationship between pulse duration and the pulse numbers required for `firing'.
For this, we denote a firing event when the current exceeds 1 \textmu A under continuous voltage pulses.
The firing ratio is defined as the percentage of such firing events for a train of 100 pulses.
We see that as the voltage pulse width widens with a fixed pulse period, the memristor more readily reaches the threshold for firing. 
This ability of the memristor to adjust its firing threshold this way underscores its potential for efficient and adaptive neuromorphic computing.

While neurons follow a threshold-based firing model, synapses facilitate information transfer by fine-tuning their synaptic plasticity in a highly parallel manner. 
The diffusion dynamics of Ag  resembles the behaviour of synaptic Ca$^{2+}$ in biological systems (Fig. 3d).\cite{ref41, ref42}
Applying large voltage pulses can induce the growth of CFs, likely because the high-rate injection of Ag$^{+}$ into the filament volume predominates over filament self-dissolution.\cite{ref17}
This process is akin to synaptic strength enhancement caused by a Ca$^{2+}$ surge in biological systems. 
Therefore, the steady-state evolution of the filament can effectively emulate synaptic plasticity.\cite{ref17}
When exposed to a large voltage pulse train (15 ms, 2.5 V; 5 ms, 0.1 V), a progressive increase in conductance is observed (Fig. 3e). 
It mirrors the continuous growth in synaptic weight between neurons, a key process in biological memory formation and strengthening. 
The synaptic weight could also be adjusted through pulse frequency and amplitude (Fig. S17). 
This feature resembles the neurological system's mechanism, where the strength of memory consolidation is influenced by the frequency and intensity of neural stimuli.\cite{ref43} 
At this stage, the current under the read voltage remains intact.
Therefore the device remains in a volatile state, exhibiting short-term plasticity (STP). 
When the voltage pulse is increased to 3.5 V, the current surges sharply, transitioning into a non-volatile state (Fig. 3f).
This is likely due to the sudden formation of robust CFs under substantial stimulus.
Within the same pulse cycle, although the synaptic current reaches a CC of 5mA during the pulse train, a gradual increase in conductance is observed under the read voltage.
The increase in current exhibits a near-linear trend (inset of Fig. 3f), which is favorable for constructing artificial neural networks (ANN).\cite{ref44} 
After the completion of the pulse train, the device continues to maintain its current under the read voltage, confirming the transition from STP to long-term plasticity (LTP), signifying the establishment of a long-term connection between the synapses. 
Collectively, by modulating the voltage pulse parameters, the device can therefore simulate the functionalities of both synapses and neurons.

\begin{figure}[H]
\centering
\includegraphics[width=\textwidth]{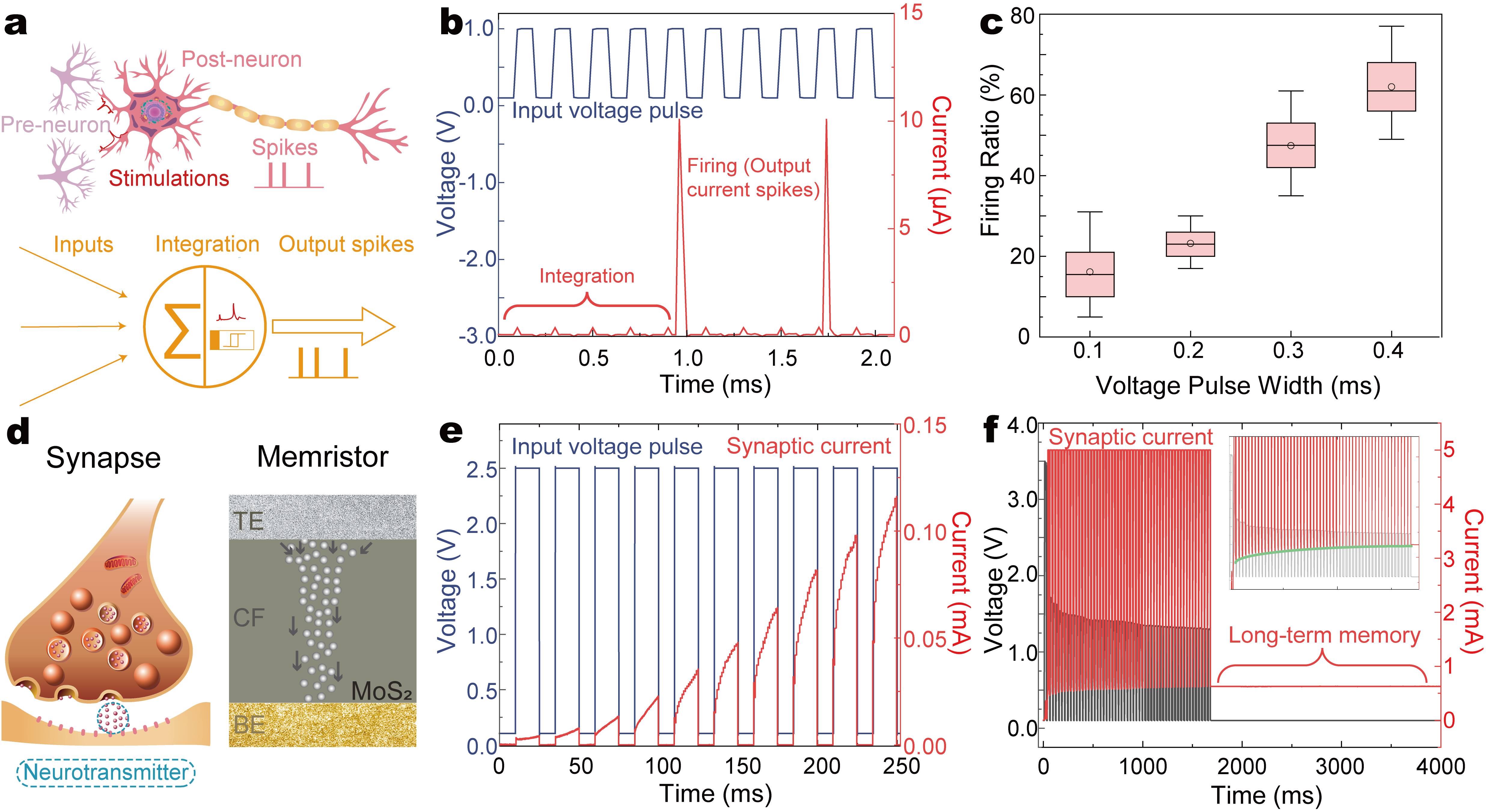}
\caption{\label{fig:3}
\textbf{Reconfigurable memristor-based neuron-like LIF functionality and synaptic plasticity.}
\textbf{a} Schematic representation of information transmission in neurons and the LIF function. 
\textbf{b} Measurement demonstrating the LIF functionality of the device operating in a volatile state (pulse train: 0.1 ms, 1 V; 0.1 ms, 0.1 V).  
\textbf{c} Box plot showcasing the dependence of the firing ratio on input pulse duration. Pulse widths are set at 0.1, 0.2, 0.3, and 0.4 ms, respectively, with an amplitude of 1.0 V and an OFF time interval of 0.1 ms. Each test includes 10 samples on same device with a sample size of 100 pulses. The hollow circle represents the mean value. 
\textbf{d} Illustration highlighting the similarity between the establishment of connections via neurotransmitters in synapses and the formation and thickening of CFs in memristors. TE, CF, and BE represent the top electrode, conductive filament, and bottom electrode, respectively. The arrows depict the process of Ag$^{+}$ ions forming the CFs.
\textbf{e} Device demonstrating short-term plasticity (pulse train: 15 ms, 2.5 V; 10 ms, 0.1 V). 
\textbf{f} Device demonstrating conductance changes under non-volatile state (pulse train: 20 ms, 3.5 V; 10 ms, 0.1 V). The inset shows a magnified view of the increasing current under the applied read voltage.
}
\end{figure}

\section{Reconfigurable memristor-based smart packaging} 

The use of paper substrate allows our printed memristors to be used for innovative smart packaging without the use of traditional electronics. 
As an example, we physically implement a memristor-based tri-mode electronic Time-Temperature Indicator (TTI). TTI is a crucial component in smart packaging\cite{ref45} for applications demanding strict temperature regulation, such as food and pharmaceutical industries (Fig. 4a) \cite{ref46, ref47}. 
It dynamically tracks ambient temperature changes, integrates these variations to compile thermal history, and identifies exposures to abnormal temperatures that could compromise product quality. 
Compared to traditional TTIs, such as those based on chemical or enzymatic reactions\cite{ref48, ref49}, electronic TTIs offer real-time, continuous temperature monitoring with enhanced accuracy and reliability.
However, manufacturing complexity and cost limit the wide adoption of electronic TTIs.\cite{ref50}

The operation principle and conceptual design of the electronic TTI are illustrated in Figs 4b and 4c, respectively.
Specifically, reconfigurable memristors simplify the neuronal circuit design (Fig. S18) and concurrently function as both the processor and data storage unit (Fig. 4b). 
Its LIF functionality tracks thermal history of the package during its journey through the supply chain while providing over-threshold warning with the reconfigurable properties of its non-volatile state. 
To test our TTI implementation, we use different voltage pulses to represent temperature sensor output corresponding to different temperatures.
The memristor efficiently processes these signals and converts them into a time-temperature integral. 
Its `leaky' feature only accounts for information reaching the firing threshold, effectively filtering out insignificant temperature fluctuations. 
In contrast, significant temperature variations are recognized by the device transitioning into a non-volatile state, thereby maintaining the information.
Figure 4d illustrates the implementation of our TTI system. 
Using a simple indicator like a microLED, the output from the TTI can be displayed (Fig. S19, Supporting movies). 
When the temperature remains within the ideal range, the time-temperature integral is insufficient to trigger any firing event (Fig. 4e). 
The LED remains off, indicating the product quality is assured. 
Upon exposure to moderately elevated temperatures, the system commences the integration of temperature signals.
As thermal conditions persist, the time-temperature integral surpasses the threshold, triggering the firing events.
Consequently, the LED would flash, serving as a warning to cold chain management about transient deviations (Fig. 4f).
Once the temperature returns to the normal range, the display system reverts to the off state due to the memristor’s `leaky' characteristic. 
This suggests that the memristor-based TTI system is capable of effectively conducting real-time monitoring of minor and short-lived temperature anomalies.
Furthermore, significant temperature exceedance over an extended period or extreme fluctuations in a brief period would lead to the formation of robust CFs (as demonstrated in Fig. 3f), pushing the time-temperature integral far beyond the threshold.
Under such circumstances, robust CFs form, shifting the memristor into a non-volatile state (Fig. 4g).
The resulting continuous illumination of the LED signals potential spoilage of the packaged contents.

By replacing the gold (Au) bottom electrode with an inkjet-printed graphene electrode, the threshold for initiating firing events in the device can be further lowered to 0.4 V (Fig. S20). 
This underscores not only the potential for further reduction in energy consumption but also possibilities in broadening and refining the range of operating temperatures, and extending application scopes by replacing materials in inkjet printed memristors.
Additionally, the memristors can be coated with parylene to obtain a water proof TTI operation, highly desirable for real-world applications (Fig. S21). 
The TTIs based on printed reconfigurable memristors thus have potential in disposable and smart packaging.

\begin{figure}[H]
\centering
\includegraphics[width=\textwidth]{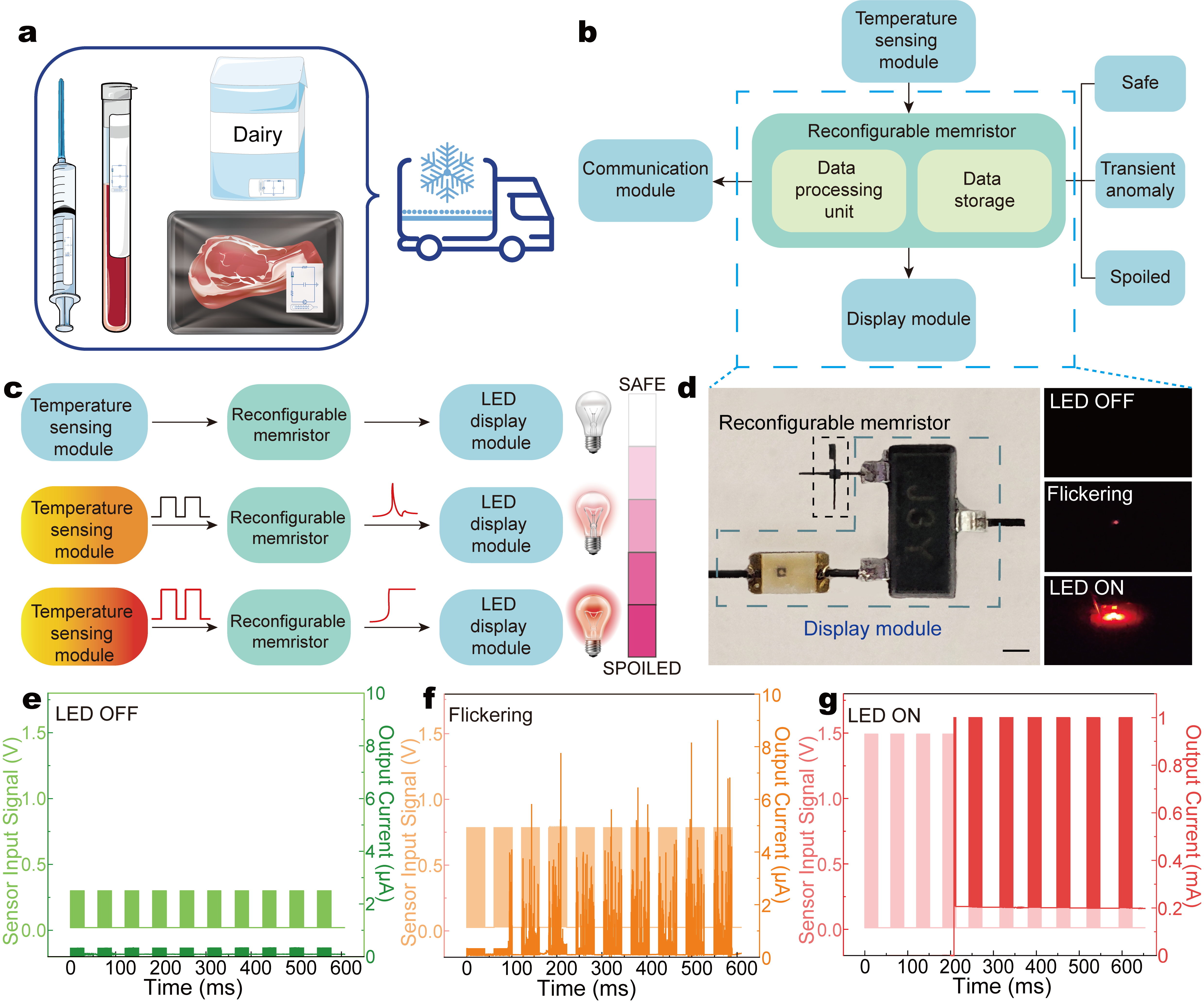}
\caption{\label{fig:4}
\textbf{Paper-based memristors for smart packaging application.}
\textbf{a} Illustration of the role of TTI for cold chain distribution.
\textbf{b} The operation mechanism of the electronic TTI where the memristor can serve both as the data processing and data storage unit.
\textbf{c} Conceptual design of the TTI system based on our reconfigurable memristor.
\textbf{d} The display and output of tri-mode TTI system in three different modes. The scale bar represents 500 \textmu m.
\textbf{e-g} Signal outputs from the memristor based on varying temperature sensor inputs with a 100x10 pulse train: 
\textbf{e} The applied pulses are 0.2 ms-0.3 V and 0.1 ms-0.1 V. The current response indicates no temperature anomaly.
\textbf{f} The applied pulses are 0.2 ms-0.8 V and 0.1 ms-0.1 V. The current response indicates minor and short-lived temperature anomaly.
\textbf{g} The applied pulses are 0.2 ms-1.5 V and 0.1 ms-0.1 V. The current response indicates significant temperature exceedance.
}
\end{figure}

\section{Memristor array for edge computing in medical image analysis}

The multilevel conductance states that can be implemented with our paper-based memristor system may provide artificial neural networks (ANNs) with the capability for refined weight adjustments, potentially enhancing ANN-based image recognition performance. 
Moreover, its lightweight neural networks for edge computing facilitate local processing, which not only minimizes cloud data transfer but also amplifies data security and transmission speeds.
This approach offers distinctive value for medical image analysis at the point-of-care. 
The democratization of early detection for 'stealth diseases' necessitates platforms that are user-friendly and economically efficient.
Our system not only offers portability and cost-effectiveness but may also allow disposal of physical implementations after use with minimal environmental footprint.

Here we propose a memristor-based DR screening system. 
This incorporates a designed algorithm for feature extraction\cite{ref51} and utilizes memristor array-based $81\times16\times2$ ANN for subsequent feature analysis.
Each memristor functions as a synapse connecting neurons across consecutive layers, with its conductance states representing the synaptic weight.
The neurons are defined by a mathematical model that incorporates a weighted sum followed by an activation function.

DR is a severe complication affecting approximately one-third of diabetic patients and is the primary cause of blindness within this demographic.\cite{ref52} 
The stealthy progression of DR often results in significant retinal damage before being diagnosed, owing to the lack of symptoms in its early stages.\cite{ref53,ref54} 
Our DR screening system conducts lesion analysis offline by edge computing to identify abnormal changes in retinal blood vessels. 
In contrast to existing DR testing software that requires uploading raw images to the cloud for analysis, our system only reports post-analysis grading results, without the need for data transfer (Fig. 5a).

In our approach, retinal images with retinopathy are imported from the DiaretDB1 dataset, followed by feature extraction. \cite{ref55} 
In the raw image (Fig. 5b), the backgrounds containing the optic disc and vasculature are first removed to enhance the visibility of candidate lesions (Fig. 5c). 
The algorithm then successively performs binarization and morphological operations on the images to remove noise and identify the candidate lesions.
Based on distinct appearance and underlying pathogenesis, lesions are categorized into bright lesions (Fig. 5d) and red lesions (Fig. 5e). 
Red lesions primarily comprise retinal hemorrhages and microaneurysms, while bright lesions mainly encompass hard exudates and soft exudates. \cite{ref52} 
For each type, the algorithm separately extracts the structural, color, and derivative features and converts these features into 81-dimensional vectors (Fig. S22).
The vectors are subsequently input into a designed memristor array based $81\times16\times2$ ANN for feature analysis. 
We use the synaptic plasticity of the memristors to conduct simulation. 
Considering the variance of our printed devices, we randomly select 9 devices in the 30-layer memristor array in Fig. 2a and input their conductance potentiation and depression cycles (Fig. S23). 
During a forward pass, the processed data is fed into the ANN, yielding a grading result. 
The ANN undergoes training through the back propagation algorithm, differentiating regions with lesions from those without and adjusting synaptic weights according to output errors.
This training encompasses 1000 epochs.
The detection result for the representative image is shown in Fig. 5f, with the lesion areas marked in blue. 
Fig. 5g summarizes the overall accuracy and specificity for all four lesion types, each exceeding 90\%.
These simulated results highlight the promising capabilities of our printed memristors for potential applications in transient or disposable medical image processing.

\begin{figure}[H]

\includegraphics[width=0.786\textwidth]{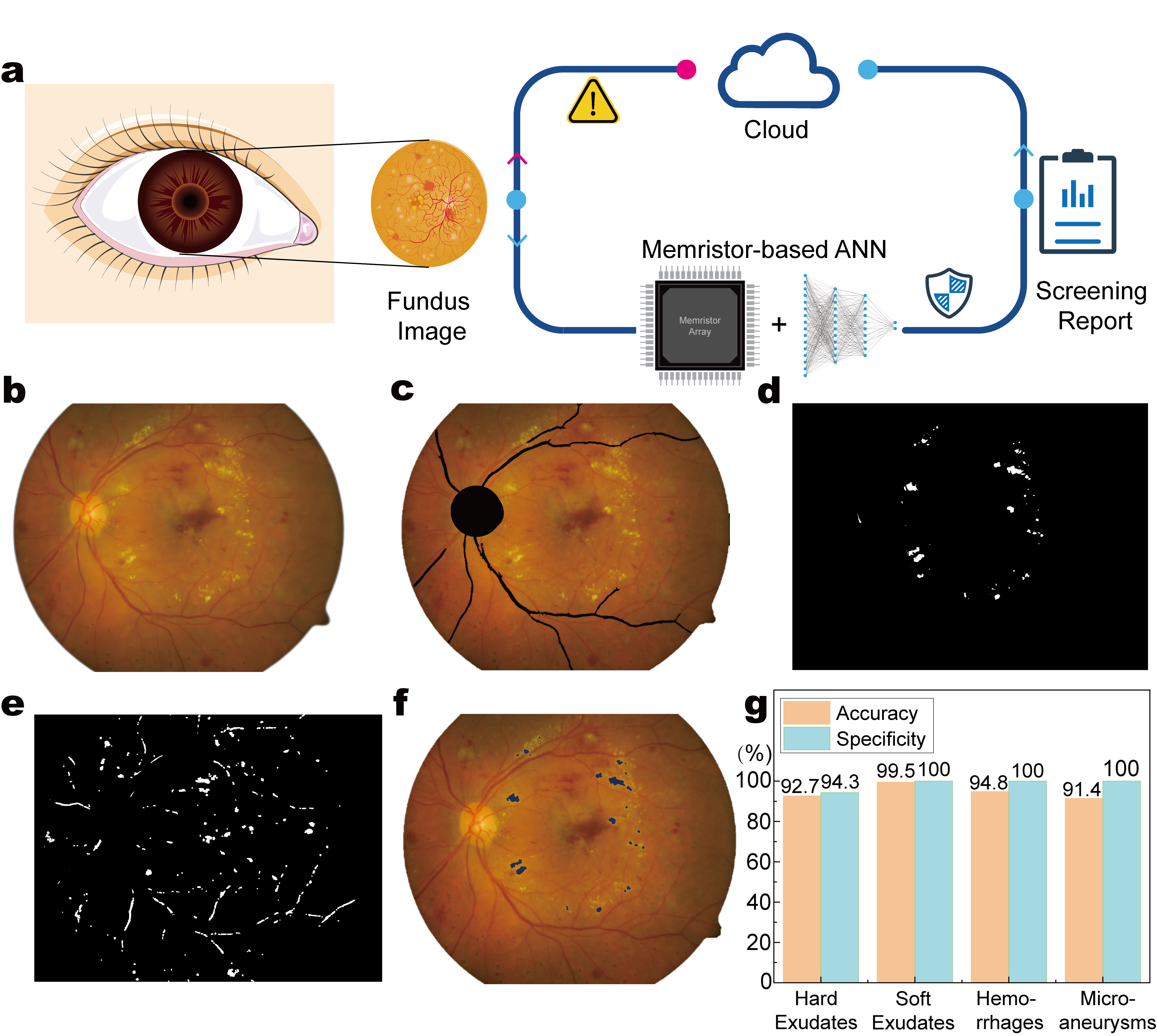}
\centering
\caption{\label{fig:5}
\textbf{Paper-based memristors for sustainable image processing-based POCT applications}
\textbf{a} Illustration of an edge computing system designed for offline DR screening.
\textbf{b} Original retinal image sourced from the DiaretDB1 dataset.
\textbf{c} Processed image with both the optic disc and vasculature removed.
\textbf{d} Bright lesion candidates after feature extraction.
\textbf{e} Red lesion candidates after feature extraction.
\textbf{f} Lesion recognition outcomes using the developed ANN; blue dots pinpoint the lesion locations.
\textbf{g} Performance evaluation of the memristor system in the task of DR image recognition.
}
\end{figure}

\section{Discussion}
Our paper-based, inkjet printed, reconfigurable memristors exhibit outstanding resistive switching performance.
The recyclability of the memristors emphasizes an important step towards sustainable electronics.
By demonstrating our memristors in smart packaging and designed framework for medical image processing, we underscore the potential of our paper-based devices. 
The technology barrier to implement our approach for biodegradadblee polymers is likely to be minimal.
Our research therefore provides valuable insights into the development of sustainable and efficient neuromorphic systems and opens up new avenues for future paper and potentially, other biodegradable substrate-based electronics.

\section{Methods}

\subsection{HPH process for MoS$_2$ ink preparation}
The PSI-40 high-pressure homogenizer, together with the D202D diamond interaction chamber (dual slot deagglomeration chamber with microchannel dimension of 87 \textmu m), is used to make printable ink of MoS$_2$ nanoflakes. 
MoS$_2$ powder ($\sim$6 \textmu m, from Sigma-Aldrich) is first mixed with anhydrous IPA (from Sigma-Aldrich) at a concentration of 5 mg mL$^{-1}$ to formulate the raw dispersion. 
During the high-pressure homogenization (HPH) process, a combination of cavitation, shear, and impact forces reduces the size of particles and exfoliates the MoS$_2$ crystals into nanoflakes. 
This nanoflake dispersion is then collected and centrifuged for 1 hour at 4000 rpm (corresponds to a relative centrifugal force of 1520$g$) with a Hettich Universal 320 Benchtop Centrifuge. 
This centrifugation step separates the supernatant containing exfoliated MoS$_2$ nanoflakes from the unexfoliated materials (sediments within the container after centrifugation). 
The supernatant is collected as the raw IPA ink. 
This concetration of MoS$_2$ is adjusted for inkjet printing, with 2-butanol (from Sigma-Aldrich) added into the ink in a volume ratio of 10\% to reduce the coffee ring effect of the printing process by inducing Marangoni flow. 
The final concentration of the MoS$_2$ ink is 4 mg mL$^{-1}$.
The graphene ink is also formulated using HPH technique, with the addition of 3\% polyvinylpyrrolidone (PVP). The concentration of the graphene ink is 2 mg mL$^{-1}$.

\subsection{UV-Vis measurements}
A Cary 7000 UV-VIS-NIR Spectrometer is used to measure the absorbance of the MoS$_2$ ink at a given concentration. 
The Lambert-Beer law is used to calculate the concentration of the ink based on the measured absorbance, the optical path length (10 mm) of the PMMA cuvettes, and the extinction coefficient of MoS$_2$ nanoflakes at 672 nm (3400 Lg$^{-1}$ m$^{-1}$).\cite{ref56}

\subsection{AFM measurements}
Bare silicon wafers are used as substrates for the deposition of MoS$_2$ nanoflakes for AFM characterization. 
Substrates are cleaned by sonication in acetone and IPA (5 mins for each step) and are rinsed with DI water multiple times before drying under nitrogen blow. 
Then 1 ml (3-aminopropyl)triethoxysilane (APTES) is added to 50 mL DI water to make the APTES solution. 
The silicon substrates are left in the APTES solution for 15 mins to form a self-assembled monolayer (SAM) of APTES. 
The substrates are then put into clean water to stop the self-assembling process and get thoroughly rinsed. 
After this, the substrates are dried again with nitrogen blow. 
Diluted MoS$_2$ solutions (0.005 gL$^{-1}$) are drop-cast onto these APTES-modified substrates and left for 40s to let MoS$_2$ nanoflakes bind with the APTES monolayer.
The samples are then rinsed thoroughly with DI water to remove the unbound flakes. 
Then samples are finally dried and measured with a Bruker Icon AFM to get statistics of the lateral size and thickness of the flakes.

The CAFM measurements are conducted in a Bruker Dimension Icon Pro Instrument, using a Bruker Extended TUNA module. 
For the measurements, we use conductive Pt-Ir coated soft SCM-PIC-V2 tips (Bruker make, spring constant = 0.1 N/m, nominal tip radius 25 nm).

\subsection{Device Fabrication}
PEL60 printing is used as the substrate. A 5 nm layer of chromium (as an adhesion layer) and a 20 nm layer of gold are deposited sequentially on the substrate using an Edwards E306A Thermal Evaporator to form the bottom electrode. 
The evaporation process is maintained at low pressure (2$\times$10$^{-6}$ mbar) and a low rate (0.1 Å/s). 
A Fujifilm Dimatix DMP-2800 inkjet printer is used for inkjet printing. 

The 2D MoS$_2$ and graphene inks are printed with a drop spacing of 25 \textmu m.
The Ag ink (from Sigma-Aldrich) is printed perpendicular to the bottom electrodes with a drop spacing of 25 \textmu m and is then annealed at 100 $^\circ$C for 1 hour in the glovebox to form the top electrodes.

\subsection{Device level electrical characterization}
Devices are characterized using a Suss MicroTec Probe Station connected to a 4200-SCS Keithley semiconductor analyser and B2902A source measure unit (SMU).

\subsection{XPS measurements}
Inkjet printed MoS$_2$ samples are sent to the Harwell XPS Center for characterization.

\subsection{BDT Treatment Procedure}
MoS$_2$ layers are deposited on a paper substrate with evaporated gold electrodes via inkjet printing. The samples are then transferred into a nitrogen-filled glove box. Inside the glove box, a 50 mM BDT solution in anhydrous hexane is prepared, and the samples are gently immersed in this solution. The container is sealed and left to sit for 24 hours. Afterwards, the samples are soaked and gently rinsed in anhydrous hexane. Subsequently, the samples are annealed at 75 \textdegree{}C for an hour within the nitrogen glove box. Finally, silver electrodes are printed

\subsection{Device Recycling Process}
The memristors that are printed on PEL60 printing paper are subjected to bath sonication in 5 ml of Triethylene Glycol Monomethyl Ether for Ag electrode removal. 
Following this, the paper is removed, gently rinsed with IPA, and allowed to dry. 
The Triethylene Glycol Monomethyl Ether-Ag nanoparticle dispersion produced from this process then undergoes  vacuum filtration, with the filtrate, containing the silver nanoparticles, being collected and subjected to a 15-minute bath sonication.

Bath sonication in 5 ml of IPA is performed on the dried PEL60 printing paper to detach the inkjet-printed MoS$_2$. 
The paper is then removed and rinsed with IPA in preparation for re-printing.
The filter paper from the first vacuum filtration step is also immersed in IPA and subjected to ultrasonic treatment to recover any MoS$_2$ flakes that may have detached along with the Ag electrode and been retained by the filter paper.
The resulting dispersion is subjected to solvent exchange via high-speed centrifugation for the dispersion of MoS$_2$ in NMP. 
Following this, a second vacuum filtration step is performed. 
The precipitate collected on the filter paper, which consists of MoS$_2$ flakes, is then immersed in IPA and subjected to a 15-minute bath sonication. 

\subsection{Simulation of DR screening}
To leverage the difference between the conductance values of synaptic devices, errors between the output vector and ground truth are back-propagated to each layer during the weight updating process. 
Here, weights are either potentiated or depressed according to the output of a sign function. 
A three-layer multi-layer-perceptron ($81\times16\times2$) is constructed and trained on the DiaretDB1 dataset, following a three-stage system flow with parametric changes and a confidence level of 75\%. 
A multi-layer perceptron (MLP) is an advanced type of neural network that consists of multiple layers of perceptrons, each layer connected to the next.
It is capable of learning complex functions by having multiple layers of neurons, each applying nonlinear transformations to the input data.
Initially, removal of backgrounds comprising optic disc and vasculature from the fundus images is conducted. 
Subsequently, a feature extraction phase is implemented, wherein the structural, color, and derivative attributes of candidate lesions are calculated and transformed into 81-dimensional vectors. 
Lastly, the entire system is trained on these vectors over a course of 1000 epochs.

\section{Acknowledgements}
M.X., N.M. and T.H. acknowledge support from Engineering and Physical Sciences Research Council (EP/T014601/1, EP/L016087/1). G. Y. acknowledges support from Royal Society. B.Z., F.T. and Z.C. acknowledges support from China Scholarship Council.

\section{Author information}
Jinrui Chen and Mingfei Xiao: These authors contributed equally to this work.

\subsection{Contributions}
J.C, M.X., and K. W. fabricated the devices. J.C. and K.W. performed device characterization. M.X. and N.M. conducted ink formulation and characterization. Z.C., S.K., and J.C. designed the DR algorithm. Z.C, B.Z., G.Y. and Y. X. contributed to the discussion of results. B.Z. conducted the BDT-treatment experiment. S.F. conducted the cross-sectional TEM imaging. R.O., S.G. and J.C. designed and conducted the CAFM measurement experiments. T.H. directed and coordinated the research. J.C., M.X., Z.C. and T.H. wrote the manuscript. All authors participated in the scientific discussion and contributed to the writing of the manuscript.

\subsection{Corresponding author}
Correspondence to Tawfique Hasan (th270@cam.ac.uk)
\end{document}